\documentclass[aps,preprint,amsmath,amssymb]{revtex4}
\usepackage{graphicx}
\begin{document}
\title{Enhancement of equilibrium fraction by heterogeneous nucleation in a ``phase separated'' half-doped manganite.}
\author{P. Chaddah, Kranti Kumar and A. Banerjee} 
\affiliation{UGC-DAE Consortium for Scientific Research\\University Campus, Khandwa Road\\
Indore-452017, M.P, India.}
\date{\today}
\begin{abstract}
Glass-like arrest of kinetics has been observed across many magnetic first-order transitions. By traversing the two control variable H-T space, tunable coexisting fractions of arrested and equilibrium phases have been observed. We report here a fortuitous situation in a half-doped manganite sample, where these features occur by varying only temperature (T) along the H=0 line. The resistivity at 5 K rises by more than a factor of three provided second cool-down is effected from a specified intermediate T. This significant enhancement results from heterogeneous nucleation during second cool-down of regions that were kinetically arrested during first cool-down.
\end{abstract}
\maketitle
    Rapid cool down, or splat-cooling, is used to form metallic glasses since the kinetics can then be arrested before the low-temperature structure can be formed. Such an arrest of kinetics can also inhibit a 1st order transition where both the phases, on either side of the transition, have long range structural (including magnetic) order. The high-T phase persists in the low-T region, where it is energetically unstable, and the lack of dynamics triumphs over thermodynamics. Magnetic 1st order transitions, in a variety of magnetic materials, have recently been shown to exhibit such a kinetic arrest \cite{man, kjs,chat,chat2,chad,kran,roy1,ban1,ban2,roy2,rawat,roy3,sharma,chad2,pk}. The transition in these materials is influenced strongly by the magnetic field applied. Many functional materials are multicomponent systems with intrinsic disorder, and their 1st order transitions occur across a broadened band (rather than across a sharp line) in H-T space. These bands consist of lines corresponding to regions of length-scale of the order of the correlation length, and this has been used imaginatively to create controlled fractions of coexisting unstable (but kinetically arrested) and equilibrium phases \cite{kran, ban2,rawat}.

    The free energy has two equal minima, separated by a barrier, at T=Tc. The high-T  phase, however, can persist till a lower T=T* where this barrier vanishes  \cite{chad3} and we refer to this limit as the supercooling spinodal. Homogeneous nucleation of the low-T phase is expected only close to T*. The transition could be caused above the spinodal either by introducing energy fluctuations \cite{chad3,roy4}, or by heterogeneous nucleation. In this paper we consider a half-doped manganite where homogeneous nucleation is inhibited in some regions because the T$_K$ of these regions lies above the corresponding T*, but we show that heterogeneous nucleation of the antiferromagnetic insulating phase can be caused at a higher temperature. 

    Many half-doped manganites show a ferromagnetic-metallic (FMM) to antiferromagnetic-insulating (AFI) transition as T is lowered, but cooling in a large field results in the FMM phase being kinetically arrested and persisting to low-T \cite{ban2,rawat,chad2}. By choosing an appropriate H for the cooling process, the frozen FMM fraction can be tuned. It was conjectured that for some samples corresponding to figure 5(c) of ref.\cite{chad2}, this coexistence of an unstable FMM phase would be seen even on cooling in zero field. We present here results for one such half-doped manganite (La-Ca-Mn-O) which shows ``phase separation'' in zero field \cite{ban3}. As has been found for various materials earlier \cite{kran,ban2,roy2,rawat,pk}, the T* and T$_K$ are anticorrelated in this sample also. This implies that those regions (see above) which have a higher value of T$_C$ or of the supercooling spinodal T*, have a lower value of the kinetic arrest temperature T$_K$ \cite{chad}. This is depicted in figure 1, which draws on the similar figures in refs. \cite{chad,kran,roy2}. 

    As the sample is cooled in zero H towards point A, we have a homogeneous FMM phase even though it is metastable. At point A, the entire sample is above the supercooling spinodal and no homogeneous nucleation takes place. Since regions corresponding to W-band get kinetically arrested at point B, before there is any homogeneous nucleation, these regions remain FMM at the lowest temperature of point C. But regions corresponding to bands X,Y and Z have got converted to AFI at this point C since their respective spinodal T* is higher than their corresponding T$_K$. We now warm the sample to point B. We retain X, Y and Z in AFI phase, but W is kinetically arrested. We now warm toward point A. W-band is no longer kinetically arrested, is above its supercooling spinodal, but is sitting in an environ where many nuclei of AFI phase exist corresponding to bands X, Y and Z. There is now a possibility of  FMM phase of band W undergoing heterogeneous nucleation and converting to the AFI phase. We show through resistivity measurements that such a heterogeneous nucleation actually takes place. If the fraction in W band is just around percolation threshold, then its conversion from FMM to AFI results in a sharp rise in resistivity. 

Figure-2 shows the resistivity measurement in zero field. The inset shows the thermal hysteresis across the first-order FMM to AFI phase while cooling from 320 K to 5 K and again heating from 5 K. However, in this cooling process, the complete transformation to the AFI state has not taken place even after approaching the lowest temperature. This is evident from the main panel of figure-2 where instead of heating all the way from 5K to 320 K the sample is heated up to 150 K and cooled back again to 5K. A spectacular increase in resistivity takes place, giving rise to more than three times increase in resistivity at the lowest temperature.  The temperature 150 K was chosen based on detailed data obtained by spanning the H-T space \cite{ban3}.  As discussed above, we attribute this rise in resistivity to the increase in the AFI phase fraction. This additional AFI phase was formed by heterogeneous nucleation.   

In this report we have studied the fortuitous situation envisaged in ref. \cite{chad2}, where cooling even in H=0 gives a coexistence of glassy FMM with equilibrium AFI phase at low-T. While a homogeneous (but kinetically arrested) FMM phase can be obtained by cooling in a large H, a homogeneous AFI phase appears not realizable. By the process of heating to just above the T$_K$ band, we have invoked heterogeneous nucleation and enhanced the equilibrium AFI phase. The de-arrest just above T$_K$ used here corresponded to the ``softening'' of glass. Cooling in high field traps a much larger fraction of FMM glass, and warming in zero field would cause regions Z, Y,.. to cross T$_K$ while still below the corresponding T*. The de-arrest would now transform glassy FMM to AFI, and these measurements will report \cite{ban3} analogies to ``shattering'' of glass.    

We thank S. B. Roy for discussion. DST, Government of India is acknowledged for funding 14 Tesla PPMS.

\begin{figure}
	\centering
		\includegraphics{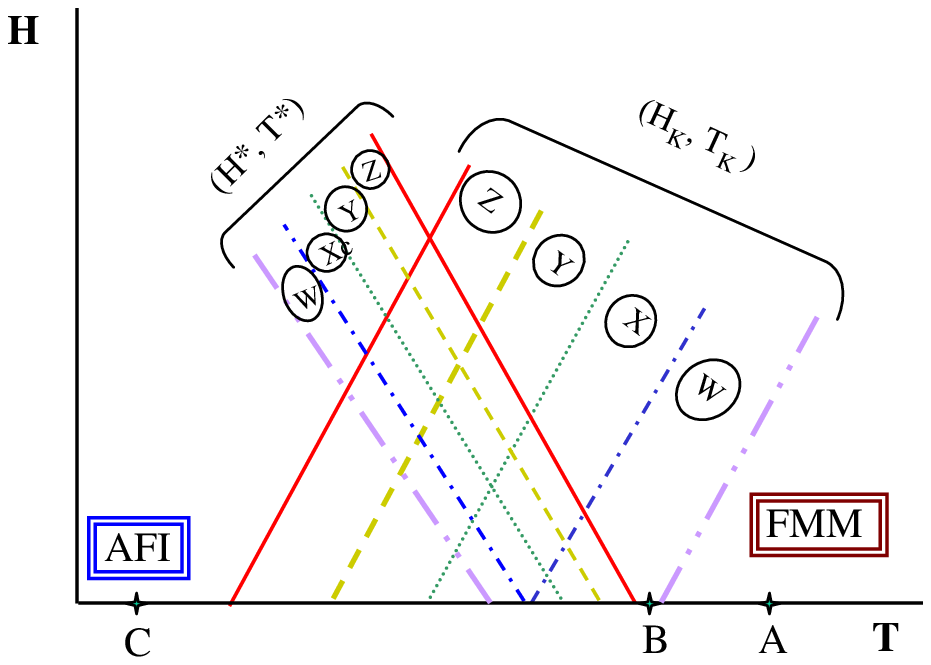}
	\caption{The high-T phase is FMM and the low-T phase is AFI. The phase transition would occur in a (H$_C$, T$_C$) band, but the FMM phase can be supercooled to the (H*, T*) band. Glass-like arrest of kinetics will occur at (H$_K$, T$_K$) band. Anticorrelation between supercooling and kinetic arrest is assumed (see ref. \cite{chad,kran, roy2} for details). Points A, B and C are as referred to in text. }
	\label{fig:Fig1}
\end{figure}

\begin{figure}
	\centering
		\includegraphics{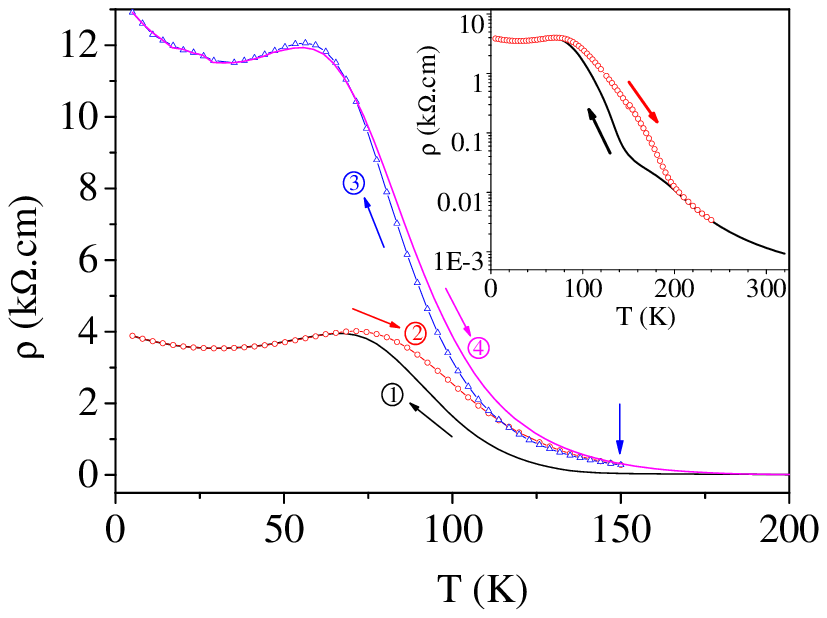}
	\caption{Resistivity vs. T is measured in zero field. Inset shows complete thermal cycling between 320 and 5 K bringing out the first-order transition. Main panel shows the result of intermediate cool-down from 150 K in paths 3 and 4. The sharply rising resistivity is a consequence of additional AFI phase formed by heterogeneous nucleation. }
	\label{fig:Fig2}
\end{figure}

\end{document}